\documentclass[aps,floatfix,showpacs,twocolumn]{revtex4}
\usepackage{amsmath,amssymb,graphicx}

\begin{document}
\title{Hierarchical solutions of the Sherrington-Kirkpatrick
model: Exact asymptotic behavior near the critical temperature}

 \author{V.  Jani\v{s}} \author{A. Kl\'\i\v c}

\affiliation{Institute of Physics, Academy of Sciences of the Czech
Republic, Na Slovance 2, CZ-18221 Praha, Czech Republic }

\date{\today}

\begin{abstract}
  We analyze the replica-symmetry-breaking construction in the
  Sherrington-Kirkpatrick model of a spin glass. We present a general
  scheme for deriving an exact asymptotic behavior near the critical
  temperature of the solution with an arbitrary number of discrete
  hierarchies of the broken replica symmetry.  We show that all
  solutions with finite-many hierarchies are unstable and only the
  scheme with infinite-many hierarchies becomes marginally stable. We
  show how the solutions from the discrete replica-symmetry-breaking
  scheme go over to the continuous one with increasing the number of
  hierarchies.
\end{abstract}
\maketitle

\section{Introduction}
\label{sec:Intro}

Spin glasses are specific unusual statistical systems, since except for a
few limiting cases no exact analytic solutions are available, even for
mean-field models. A formal exact solution of the mean-field Ising spin
glass, Sherrington-Kirkpatrick (SK) model, is known in form of the
replica-symmetry-breaking (RSB) scheme of Parisi.\cite{Parisi80a} The
solution of the RSB construction is, however, explicitly known only
approximately. One can  iterate the RSB scheme with finite many
hierarchies. In this way 1RSB and 2RSB solutions were explicitly
calculated.\cite{Parisi80b} Or one can go over to the limit of infinite
number of hierarchies and try to find a solution to the continuous limit
of the RSB scheme controlled by a nonlinear partial differential equation.
A solution to the continuous RSB scheme is known near the critical
temperature.\cite{Parisi79,Parisi80a,DeDominicis83,Kondor83,Sommers85}
Recently, a numerical solution of the differential equation from the
continuous RSB scheme was obtained also beyond the critical region at zero
magnetic field.\cite{Crisanti02}

The RSB construction has hence presently two analytically accessible
limits: the discrete scheme with a few hierarchies and the continuous
limit. The former, within 1RSB or 2RSB schemes, can be applied
anywhere in the low-temperature phase. The latter, on the other hand,
is viable practically only near the critical temperature at zero
magnetic field.  Originally the discrete scheme was considered as an
approximation to the ultimate solution, the continuous limit.
Recently, however, variants of mean-field spin glass models such as
random-energy \cite{Derrida80}, $p$-spin \cite{Gardner85} or Potts
\cite{Elderfield83}, display regions where already 1RSB scheme becomes
stable. The discrete RSB scheme has thus won its own substantiation
beyond a mere approximate scheme. Moreover, using the concept of
thermodynamic homogeneity one can derive the discrete RSB scheme with
finite-many hierarchies without any specific assumptions on the
behavior of the order parameters.\cite{Janis05} Even in the limiting
case of the discrete RSB scheme with infinite-many hierarchies one
need not end up with the continuous limit. To derive the continuous
limit one has to assume that both differences $\Delta q_i = q_{i+1}
-q_i$ and $\Delta m_i = m_i - m_{i+1}$ from the Parisi construction
are infinitesimal (of order $1/K$) if the number of hierarchies $K$
approaches infinity. There is no \'a priori reason for such a uniform
behavior and unless proved by explicit calculations, it must be
assumed as an ansatz, as actually done by Parisi in his derivation of
the full RSB solution.

The number of hierarchies used in the RSB solution is generally a free
parameter that cannot be derived from free energy. This number is
determined from stability of the thermodynamic equilibrium state. To
decide how many hierarchies in the RSB scheme are needed to reach a
thermodynamically stable solution and whether the infinite number of
hierarchies leads to the continuous distribution of the order
parameters one has to analyze the RSB scheme with a variable number of
hierarchies.  This is possible only in specific asymptotic limits such
as the asymptotic region near the critical temperature. The existing
analyses\cite{Parisi79,Parisi80a,DeDominicis83,Kondor83,Sommers85} of
the RSB scheme near the critical temperature use an incomplete
expansion of free energy where only a single, simplest term from the
highest-order contribution is taken into consideration.  Such an
expansion does not reproduce correctly the asymptotic behavior of the
RSB solutions with a few hierarchies (1RSB, 2RSB). It hence cannot
serve as a proof of validity of the continuous limit for the RSB
solution with infinite-many hierarchies.
 
The aim of this paper is to present a general scheme for analyzing the full
discrete hierarchy of RSB solutions for mean-field models of spin glasses.
We in particular concentrate on the SK model at zero magnetic field  and
use the asymptotic solution to determine the way in which the limit of
infinite many hierarchical levels is approached.  With the exact
asymptotic solution for arbitrary numbers of RSB hierarchies we prove that
the stable solution of the SK model  is the continuous limit of the RSB
scheme. We explicitly evaluate the equations for the order parameters in
the first nontrivial order of the expansion parameter $\theta = 1 - T/T_c$
around the critical temperature $T_c$. We perform the expansion generally
for the RSB solution with $K$ hierarchies. To achieve this goal, we use an
explicit representation for the hierarchical free energy (partition
function) with $K$ levels from which we derive equations for the order
parameters. These equations are then asymptotically expanded near the
critical temperature to obtain an explicit leading-order asymptotic
behavior of all the order parameters at any hierarchical level of the RSB
construction. From it we reconstruct the continuous version of the RSB
scheme when the number of hierarchies is limited to infinity. We
explicitly show that any solution with a finite number of hierarchies is
unstable even arbitrarily close to the critical temperature  and only the
continuous limit, as expected, becomes marginally stable in the spin-glass
phase.\cite{DeDominicis83}

\section{RSB solution with $K$ hierarchies}
\label{sec:KRSB-equations}

We avoid the replica trick and use the explicit representation of the RSB
solution with generally $K$ hierarchies from Ref.~\cite{Janis05} derived
from the thermodynamic approach of Thouless, Anderson, and Palmer. In this
representation the solution has $2K + 1$ order parameters. There are $K +1$
physical ones, $q,\Delta\chi_1,\ldots, \Delta\chi_K$ related to various
ways squares of local magnetizations can be calculated. The other $K$
parameters, $m_1,\ldots, m_K$,  have a geometric origin and are connected
with the way we break the replica symmetry.

To derive an analytic representation for the $K$-level hierarchical free
energy we start with an averaged free-energy density of the
Sherrington-Kirkpatrick model with $\nu$ real replicas
\begin{multline}\label{eq:FE-averaged-finite}
f_\nu = \frac{\beta J^2}{4} \left[\frac 1\nu\sum_{a\neq b}^\nu
\left\{\left(\chi^{ab}\right)^2 + 2 q\chi^{ab}\right\} - (1 -
q)^2\right]\\ -\frac
1{\beta\nu}\!\int\limits_{-\infty}^{\infty}\frac{d\eta}{\sqrt{2\pi}} \
e^{-\eta^2/2} \ln \text{Tr}_S \exp\left\{\beta^2J^2\sum_{a < b}^\nu
\chi^{ab}S^aS^b\right. \\ \left. + \beta \bar{h}\sum_{a=1}^\nu S^a\right\}
\end{multline}
where the trace $\text{Tr}_S$ runs over the replicated Ising spins $S^a =
\pm 1$. We denoted the internal magnetic field $\bar{h} = h +
\eta\sqrt{q}$. Averaging over the fluctuating internal field $\eta$
replaces averaging over the spin exchange  $J_{ij}$ in the mean-field
solution. The averaged order parameter  at the saddle point $q =
\langle\langle S^a\rangle_T^2 \rangle_{av}$  does not depend on the
replica index due to equivalence of  replicas. Real replicas allow us to
introduce  local inter-replica susceptibilities  as order parameters being
at the saddle point $\chi^{ab} = \langle\langle S^a
S^b\rangle_T\rangle_{av} - q$. Angular brackets $\langle\ \rangle_T$
denote averaging over the configurations of replicated spins.
 
The hierarchical solution is constructed by successive applications of
the replica symmetry ansatz on the matrix of the overlap
susceptibilities We remind that $\chi^{aa}=0$. For $K=1$ we choose
$\chi^{a\neq b} = \chi_1$ and decouple the spin variables via a
Hubbard-Stratonovich transformation and a new fluctuating
replica-diagonal field $\lambda_1$. This transformation enables us to
evaluate the trace over the replicated spins explicitly.  We recover
1RSB solution of Parisi. The next hierarchy is obtained if we replace
each matrix element $\chi^{ab}$ by a matrix $\chi^{ab}_{\alpha\beta}$
. We again apply the replica symmetry ansatz on non-diagonal elements.
We have two parameters $\chi^{aa}_{\alpha\neq\beta} =\chi_1$ and
$\chi^{a\neq b}_{\alpha\beta} = \chi_2$. The replicated spins will be
decoupled with two Hubbard-Stratonovich fluctuating field $\lambda_1,
\lambda_2$.  We proceed in this hierarchical construction up to the
desired $K$-level solution.

We can conveniently represent the $K$-level free energy recursively.
We define a sequence of partition functions \pagebreak[1]
\begin{equation}\label{eq:pf-hierarchy}
Z_l =  \left[\int_{-\infty}^{\infty}\mathcal{D}\lambda_l\
Z_{l-1}^{m_l}\right]^{1/m_l}
\end{equation}
where we used an abbreviation for the Gaussian differential
$\mathcal{D}\lambda_l \equiv {\rm d}\lambda_l\
e^{-\lambda_l^2/2}/\sqrt{2\pi}$.  The initial condition reads $Z_0 =
\cosh\left[\beta\left(h + \eta\sqrt{q} + \sum_{l=1}^{K}\lambda_l
\sqrt{\Delta\chi_l}  \right)\right]$.  The averaged free energy
density with $K$ hierarchies can then be represented as \cite{Janis05}
 \begin{multline} \label{eq:mf-avfe}
f^K(q,\Delta\chi_1,\ldots,\Delta \chi_K; m_1,\ldots,m_K) = - \frac
1\beta \ln 2 \\ + \frac \beta 4 \sum_{l=1}^K m_l\Delta\chi_l\left[2\left(q
+  \sum_{i=l+1}^{K}\Delta\chi_{i}\right) + \Delta\chi_l\right]\\
-\frac\beta 4 \left(1-q -\sum_{l=1}^K \Delta\chi_l\right)^2  - \frac 1\beta
\int_{-\infty}^{\infty} \mathcal{D}\eta\ \ln  Z_K
\end{multline}
with order parameters $q, \Delta\chi_l$ and $m_l,\ l=1,\ldots,K$
to be determined from stationarity equations.

To represent the mean-field equations we introduce a set of hierarchical
density matrices in the space of fluctuating random fields $\lambda_l$. We
define $\rho_l(\eta,\lambda_K,\ldots,\lambda_{l}) = Z_{l-1}^{m_{l}}/
\left\langle Z_{l-1}^{m_{l}}\right\rangle_{\lambda_{l}}$. We further
introduce short-hand notations $t \equiv \tanh\left[\beta\left(h +
\eta\sqrt{q} + \sum_{l=1}^K \lambda_l\sqrt{\Delta\chi_l} \right)\right]$
and $\langle t\rangle_l(\eta,\lambda_K,\ldots,\lambda_{l+1}) \equiv
\langle\rho_l\ldots\langle\rho_1 t \rangle_{\lambda_1} \ldots
\rangle_{\lambda_l}$ with $\langle X(\lambda_l) \rangle_{\lambda_l} \equiv
\int_{-\infty}^{\infty}\mathcal{D}\lambda_l\ X(\lambda_l)$.

With the above definitions we can write down the stationarity
equations for the physical order parameters
\begin{subequations}\label{eq:mfeqs-physical}
\begin{align}\label{eq:mfeqs-q}
  q^K &= \langle\langle
  t\rangle_K^2\rangle_\eta\\  \label{eq:mfeqs:Delta_chi}
  \Delta\chi_l^K &= \langle\langle \langle
  t\rangle_{l-1}^2 \rangle_K\rangle_\eta -
  \langle\langle\langle t\rangle_{l}^2
  \rangle_K\rangle_\eta
\end{align}
\end{subequations}
and for the geometric ones
\begin{align}\label{eq:mfeqs-geometric}
  m_l^K &= \frac 4{\beta^2}\ \frac {\langle\langle \ln Z_{l-1}
    \rangle_K\rangle_\eta - \langle\langle \ln Z_{l}
    \rangle_K\rangle_\eta } {\langle\langle \langle t \rangle_{l-1}^2
    \rangle_K\rangle_\eta^2 - \langle\langle \langle t \rangle_{l}^2
    \rangle_K\rangle_\eta^2}\
\end{align}
where index $l=1,\ldots,K$. Note that there is a direct connection to
the order parameters of the Parisi $K$-level RSB solution. The Parisi
parameters are identified as follows $q = q^K$ and $q_l = q^K +
\sum_{i=l}^K \Delta\chi^K_i$.

It is important to stress that the hierarchical free-energy density
from Eq.~\eqref{eq:mf-avfe} is not identical with discrete $K$-level
RSB solutions used in the literature to analyze the critical region of
the spin-glass transition in the SK model
\cite{Parisi79,DeDominicis83,Kondor83,Sommers85}.  The latter
functionals can be derived from the asymptotic limit $T\nearrow T_c$
of Eq.~\eqref{eq:mf-avfe} by using $K^{-1}$ expansion to order
$O(K^{-1})$ with an ansatz $\Delta \chi_l = \Delta_l/K$ and
$m_l/m_{l+1} = 1 + \delta_l/K$. This step is in fact a discrete
version of the continuous limit studied in Sec.~\ref{sec:properties}.
The full discrete RSB free energy equivalent to Eq.~\eqref{eq:mf-avfe}
was derived in Ref.~\cite{Dotsenko01} but not used in explicit
calculations. A necessity of modifications to the $K$-level
hierarchical free energies from
Refs.~\cite{Parisi79,DeDominicis83,Kondor83,Sommers85}.  when $K$
remains finite was discussed in Ref.~\cite{Moore02}.

The hierarchical free energy, Eq.~\eqref{eq:mf-avfe}, and the
respective stationarity equations, Eq.~\eqref{eq:mfeqs-physical} and
Eq.~\eqref{eq:mfeqs-geometric}, were derived within the real-replica
approach. The real replicas simulate dependence of the mean-field
solutions on initial conditions. The RSB scheme says that the initial
configurations of the order parameters are ordered hierarchically
according to the strength with which they influence the resulting
equilibrium state. The instantaneous value of the local magnetization
is $\tanh\left[\beta\left(h + \eta\sqrt{q} + \sum_{l=1}^K
    \lambda_l\sqrt{\Delta\chi_l} \right)\right]$, where the quenched
random field $\eta$ stands for an actual configuration of spin
couplings and the annealed random fields $\lambda_l$ for initial
configurations of the spin variables from the $l$th level with $l =
1,2, \ldots, K$.  The strength with which the $l$th level influences
the final equilibrated state is $\Delta\chi_l$. The density matrices
$\rho_l$ express probability (weight) of the $l$th level initial
configurations in the final state. Hence, $\langle t \rangle_l$
denotes a local magnetization obtained after averaging over the first
$l$ levels of initial configurations. The physical order parameters
are then constructed from various possibilities to interchange making
square and averaging over the hierarchies of the initial
configurations of local magnetizations.

The number of hierarchical levels $K$ is a free parameter in the above
construction. Its physical value leading to a unique physical solution
is determined as the smallest number needed to reach a
thermodynamically stable equilibrium state independent of initial
conditions. Thermodynamic stability of a construction with $K$
hierarchies is determined from a set of $K + 1$ stability
conditions.\cite{Janis05} If
\begin{multline}\label{eq:AT-hierarchical}
  \Lambda_l = \beta^2\left\langle\left\langle \left\langle 1 - t^2
\right.\right. \right. \\ \left. \left. \left. %
  + \sum_{i=1}^{l} m_i \left(\langle t\rangle_{i-1}^2 - \langle
    t\rangle_i^2\right)\right\rangle_{l}^2
\right\rangle_K\right\rangle_\eta \ge 0 \ .
\end{multline}
are obeyed for $l=0,1,\ldots, K$, then the solution with $K$
hierarchies describes a locally stable equilibrium state. If one or
more stability conditions are broken, we have to increase the number
of hierarchies in our construction. With increasing the number of
hierarchies used one has to observe a successive suppression of
instabilities (negativeness of parameters $\Lambda_l$) when the
construction converges. Only a convergent scheme can lead to a stable
solution for some finite or, in the extreme case, infinite number of
hierarchies.

\section{Asymptotic expansion near the critical temperature}
\label{sec:Asymptotic-expansion}

The order parameters in the spin-glass phase of the general $K$-level
free energy from Eq.~\eqref{eq:mf-avfe} cannot be solved explicitly
unless we resort to solutions with only a few hierarchical levels
($K=1,2$). The only chance to analyze the behavior of the entire
hierarchical construction, that is with arbitrary numbers of
hierarchies, is to expand the solution near the critical point where
the order parameters are small.  The hierarchical solution behaves in
an external magnetic field differently from the rotationally invariant
case. The physical order parameter $q$ and the geometric order
parameters $m_l$ remain finite at the transition line and only the
differences of the overlap susceptibilities $\Delta\chi_l$ are the
genuine small parameters controlling the expansion near the critical
point. While in the case of zero magnetic field all order parameters
determining the stationarity points of free energy asymptotically
vanish at the critical temperature.  The latter case is on one hand
more degenerate than the former, but on the other hand it is more
symmetric with less expansion terms (only terms with even parity
contribute in the rotationally invariant case). In this paper we will
analyze only the rotationally invariant case, $h = 0$.

The strategy to solve the stationarity equations for the hierarchical
free energy asymptotically near the critical temperature is to expand
the partition function into powers of the small order parameters and
restrict the solution only to a functional subspace generated by a
fixed polynomial expansion. We first use sch an expansion to derive
the leading asymptotic limit of equations for the physical parameters
$q, \Delta\chi_l$. At this stage we do not need to assume smallness of
the geometric parameters. Smallness of $m_l, l=1,\ldots, K$ at zero
magnetic field will be utilized later on when deriving the asymptotic
form of mean-field equations for them.

It appears that it is sufficient to expand the stationarity equations
for $q$ and $\Delta\chi_l$ only to the third order in these
parameters. To control the expansion we introduce an auxiliary
parameter $x$ that will be set to unity at the end.  We hence write a
canonical representation for the $l$th partition sum
\begin{multline} \label{eq:Z_l-expansion}
  Z_{l-1} = a^{(l)}_{00} + a^{(l)}_{02}\left(Y_{l+1} + c_l
    \lambda_l\right)^2 x^2 + a^{(l)}_{04}\left(Y_{l+1}\right. \\
  \left.  + c_l \lambda_l\right)^4 x^4 + a^{(l)}_{06}\left(Y_{l+1} +
    c_l \lambda_l\right)^6 x^6 + x^2 \left[ a^{(l)}_{20} \right. \\
  \left. + a^{(l)}_{22}\left(Y_{l+1} + c_l \lambda_l\right)^2 x^2 +
    a^{(l)}_{24}\left(Y_{l+1} + c_l \lambda_l\right)^4 x^4 \right] \\
  + x^4 \left[a^{(l)}_{40} + a^{(l)}_{42}\left(Y_{l+1} + c_l
      \lambda_l\right)^2 x^2\right] + x^6 a^{(l)}_{60}
\end{multline}
where we denoted $Y_l = c\eta + \sum_{i=l}^K c_i \lambda_i$, $c_l =
\beta \sqrt{\Delta\chi_l}, c = \beta \sqrt{q}$.

We now insert Eq.~\eqref{eq:Z_l-expansion} into
Eq.~\eqref{eq:pf-hierarchy} and perform the integral over the
fluctuating field $\lambda_l$. The result will again be reduced to a
sixth order polynomial in $x$. Thereby we win recursive relations for
the expansion parameters. We used the program MATHEMATICA to evaluate
the coefficients. The coefficients needed in
expansion~\eqref{eq:Z_l-expansion} are listed in Appendix~A.

Using the asymptotic expansion for the partition functions $Z_l$ we
can explicitly evaluate products of the density matrices $\rho_i$. We
will need to know the products within our precision only as quadratic
polynomials in $x$. We explicitly obtain
\begin{equation}\label{eq:rho-product}
 \prod_{i=k}^l \rho_i = 1  +  \frac{x^2}2\sum_{i=k}^l m_i c_i
  \left[ 2 \lambda_i Y_{i+1}  + c_i (\lambda_i^2 -
1)\right]\ .
\end{equation}

We use Eq.~\eqref{eq:rho-product} in the defining equations for the
physical order parameters. We denote $X_l = \sum_{i=1}^{l}
\Delta\chi_i$, $Q = q + X_K$, and $M_l = \sum_{i=1}^{l} m_i
\Delta\chi_i$ and obtain ($x = 1$)
\begin{subequations}\label{eq:q-asymptotic}
\begin{equation}\label{eq:q-asymptotic1}
q = \beta^2q \left\{ 1 - 2 \beta^2 Q + \beta^4 \left[ \frac 23 q^2 + 5 Q^2
+2 M_K\right] \right\}\ .
\end{equation}
We skip the upper index $K$ at the order parameters in order not to
make the notation cumbersome The above equation can further be
rewritten to a more suitable form
\begin{multline}\label{eq:q-asymptotic2}
  q = \beta^2 q \left\{ 1 - 2 \beta^2 Q + \frac {\beta^4}3 \left[
      2X_K^2 - 4 QX_K \right.\right. \\ \left.\left. + 17 Q^2 + 6
      M_K\right] \right\}\ .
\end{multline}
\end{subequations}
We note that $Q$ has the meaning of the Edwards-Anderson parameter. It
is, however, not an independent order parameter.

Analogously we derive the asymptotic form of the defining equations
for the differences of the overlap susceptibilities. We derive the
following asymptotic equations
\begin{multline}\label{eq:chi-asymptotic}
  \Delta \chi_l = \beta^2 \Delta \chi_l \left\{ 1 - 2 \beta^2 Q +
    \beta^4
    \left[ \frac 23 \Delta \chi_l^2  + 2 (Q - X_{l-1})^2 \right.\right. \\
  \left.\left. + 5 Q^2 - (2Q - 2 X_{l-1} - m_l) \Delta\chi_l + 2
      M_{l-1} \right]\right\}\ .
\end{multline}
Notice that to determine the leading asymptotic behavior of $q$ and
$\Delta\chi_l$ we had to keep first three orders in the expansion of
the right-hand sides of the stationarity
equations~\eqref{eq:mfeqs-physical}.  Second-order term is degenerate
in these equations and we obtain from it only the leading asymptotic
behavior of the Edwards-Anderson parameter $Q$. Going to the third
order then means that we have to calculate simultaneously to the
leading asymptotic behavior of $q$ and $\Delta\chi_l$ also the
next-to-leading asymptotic coefficient of the Edwards-Anderson
parameter.

To derive the asymptotic form of equations~\eqref{eq:mfeqs-geometric}
for the geometric parameters $m_l$ is more complicated than it was to
obtain equations~\eqref{eq:q-asymptotic} and
\eqref{eq:chi-asymptotic}. For the geometric parameters we will need
all the coefficients $a^{(l)}_{ij}$ from Appendix A, but those at the
highest power $x^6$ only for $m_l = 0$, since $m_l\sim x^2$.  In fact,
we have to expand free energy to the order $x^8$, but the
highest-order terms cancel each other due to subtraction in the
numerator on the right-hand side of Eq~\eqref{eq:mfeqs-geometric}.
After some effort we arrive at
\begin{widetext}
\begin{multline}\label{eq:m-asymptotic}
  (\beta^2 - 1) \left[\Delta\chi_l + 2 Q_{l+1}\right] m_l =
  \frac{\beta^4}3 m_l \left\{\left[ 6 \Delta\chi_l^2 + 18 \Delta\chi_l
      Q_{l+1} + 12 Q_{l+1}^2+ 6 (\Delta\chi_l + 2 Q_{l+1})X_{l-1} -
      6(\Delta\chi_l +
2Q_{l+1}) M_{l-1}\right.\right. \\ \left. \left. %
- 4 (\Delta\chi_l + 3 Q_{l+1})m_l\Delta\chi_l - 6 \sum_{i=l+1}^K
m_i\Delta\chi_i (2Q_{i+1} + \Delta\chi_i)\right] - \beta^2\left[34
Q_{l+1}^3 + 81 Q_{l+1}^2\Delta\chi_l + 64 Q_{l+1}\Delta\chi_l^2 + 16
\Delta\chi_l^3 \right. \right. \\ \left. \left.  + 10(3\Delta\chi_l^2
+ 9 Q_{l+1}\Delta\chi_l + 6 Q_{l+1}^2) X_{l-1} + 3(\Delta\chi_l + 2
Q_{l+1}) \left[2 X_{l-1}^2 + 3 \sum_{i=1}^{l-1}\Delta\chi_i (2 X_{i-1}
  + \Delta\chi_i )\right]\right]\right\}\ ,
\end{multline}
\end{widetext}
where we denoted $Q_{l+1} \equiv Q - X_{l}$. Also here we had to keep
first three orders in the expansion of the right-hand side of
Eq.~\eqref{eq:mfeqs-geometric}. In the above asymptotic form we
already used the fact that also the geometric parameters at $h = 0$
are small near the critical temperature and are of the same order as
$q$ and $\Delta\chi_l$.

Equations~\eqref{eq:q-asymptotic}-\eqref{eq:m-asymptotic} define the
asymptotic behavior of the hierarchical solution near the critical
temperature at zero magnetic field. From these equations we are able
to determine the leading asymptotic behavior of all the order
parameters for an arbitrary number of hierarchies $K$ together with
the next-to-leading order of the Edwards-Anderson parameter~$Q$. All
equations are homogeneous in the respective variable, hence a trivial
solution (paramagnetic phase) is correctly included. The asymptotic
equations are cubic in the physical order parameters and quadratic in
the geometric ones. Such a system of coupled algebraic equations is
not generally solvable. We, however, show that a nontrivial solution
in the physical sector can always be found.

\section{Nontrivial solution of the asymptotic equations}
\label{sec:Asymptotic-solution}
 
We now expand the dynamical variables in a small parameter measuring
the distance from the critical point in the spin-glass phase $\theta =
1 - T/T_c$. We have to keep only the leading asymptotic order and for
the Edwards-Anderson parameter we have to add the next-to-leading
order. Not to introduce cumbersome further indexing or new parameters
we relabel in this chapter the order parameters as follow $q \to
q\theta$, $\Delta\chi_l \to \chi_l\theta$, $m_l \to m_l\theta$, $Q \to
Q\theta + Q'\theta^2$. We expand the asymptotic equations in powers of
$\theta$ and all expansion coefficients (up to $\theta^3$) from the
left-hand side must equal those from the right-hand side.

Excluding the trivial solution from
Eqs.~\eqref{eq:q-asymptotic}-\eqref{eq:m-asymptotic} the first term on
the right-hand side defines the critical temperature.  From the second
one in Eqs.~\eqref{eq:q-asymptotic} and~\eqref{eq:chi-asymptotic} we
obtain
\begin{equation}\label{eq:Q-solution}
Q = 1\ .
\end{equation}
Notice that the Edwards-Anderson parameter in the leading order is
defined exactly by the replica-symmetric solution ($K = 0$).

The first nontrivial terms contributing to the order parameters of the
hierarchical solution are the third ones. The corresponding
coefficients at $\theta^3$ of Eqs.~\eqref{eq:q-asymptotic}
and~\eqref{eq:chi-asymptotic} are
\begin{equation}\label{eq:q-expansion}
0 = 1 - 3 Q' +  X_K^2 - 2 X_K  +  3 M_K\ ,
\end{equation}
\begin{multline}\label{chi-expansion}
  0 =  - 6 Q'  + 2\chi_l^2 + 6(1 - X_{l-1})(1 - X_{l-1} - \chi_l) \\
  + m_l \chi_l + 6 M_{l-1}\ .
\end{multline}

We use the former equation to exclude the next-to-leading coefficient
$Q'$ of the Edwards-Anderson parameter, that is
\begin{equation}\label{eq:Qprime-solution}
Q' = \frac 13 \left[1 - X_K(2 - X_K) + 3 M_K\right]\ .
\end{equation}
Using this result in Eq.~\eqref{chi-expansion} we obtain a recursive
relation
 \begin{multline}\label{eq:chi-iterative}
   m_l\chi_l = \frac 23\left[ 2 + X_K(2 - X_K) + \chi_l^2 - 3 \chi_l(1
     -
X_{l-1})\right. \\ \left.  %
- 3 X_{l-1}(2 - X_{l-1})\right] - 2 \sum_{i=l+1}^K m_i\chi_i \ .
\end{multline}
An explicit solution to this set of recursive equations reads
\begin{multline}\label{eq:chi-explicit}
  m_l\chi_l = \frac 23 \sum_{i=0}^{K-l}(2 - \delta_{i,0})(-1)^i\left[
    2 +
X_K(2 - X_K) + \chi_{l+i}^2\right. \\ \left. %
- 3 \chi_{l+i}(1 - X_{l+i-1}) - 3 X_{l+i-1}(2 - X_{l+i-1})\right]\ .
\end{multline}
 
We have now to exclude the geometric parameters $m_l$ from
Eq.~\eqref{eq:chi-iterative}. To this purpose we use
Eq.~\eqref{eq:chi-explicit} in the asymptotic expansion of
Eq.~\eqref{eq:m-asymptotic}. We then obtain another set of recursive
equations
 \begin{multline}\label{eq:m-iterative}
   m_l\chi_l^2 = \frac 12 \left(2 - 2 X_{l-1}
-\chi_l\right)\left[2 X_K(2 - X_K)  \right. \\ \left. %
- \chi_l(2 - 2 X_{l-1} - \chi_l) -
X_{l-1}(4 + 7 X_{l-1})\right. \\ \left. %
+ 9 \sum_{i=1}^{l-1}\chi_i(2 X_{l-1} + \chi_i)\right] -
3\sum_{i=l+1}^Km_i\chi_i \left[2(X_i - X_l)\right. \\ \left. %
  - (\chi_i - \chi_l)\right]\ .
 \end{multline}
 Combing Eq.~\eqref{eq:chi-explicit} and Eq.~\eqref{eq:m-iterative} we
 end up with a set of equations with only the leading-order asymptotic
 coefficients at the differences of the overlap
 susceptibilities~$\chi_l$:
\begin{widetext}
\begin{multline}\label{eq:chi-final}
  \frac 43 \chi_l \sum_{i=0}^{K-l}(2 - \delta_{i,0})(-1)^i\left[ 2 +
    X_K(2 - X_K) + \chi_{l+i}^2
- 3 \chi_{l+i}(1 - X_{l+i-1}) - 3 X_{l+i-1}(2 - X_{l+i-1})\right]\\ %
= \left(2 - 2 X_{l-1} -\chi_l\right)\left[2 X_K(2 - X_K) - \chi_l(2 -
  2 X_{l-1} - \chi_l) - X_{l-1}(4 + 7 X_{l-1}) + 9
  \sum_{i=1}^{l-1}\chi_i(2 X_{l-1} + \chi_i)\right]\\ - 4
\sum_{i=l+1}^K \left[2(X_i - X_l) - (\chi_i - \chi_l)\right]
\sum_{j=0}^{K-i}(2 - \delta_{j,0})(-1)^j\left[ 2 + X_K(2 -
X_K) + \chi_{l+i}^2 - 3 \chi_{j+i}(1 - X_{j+i-1})\right. \\ \left.%
- 3 X_{j+i-1}(2 - X_{j+i-1})\right]\ .
\end{multline}
\end{widetext}
The set of equations for $\chi_l$, $l=1,2,\ldots,K$ looks rather
complicated but we can successively express the susceptibilities
$\chi_l$ via their sums $X_{l-1}$.

Putting $l = K$ in Eq.~\eqref{eq:chi-final} we obtain a simple cubic
equation for $\chi_K$:
\begin{equation}\label{eq:chi-KK}
0 = \chi_K(2 - 2 X_{K-1} - \chi_K)(2 - 2 X_{K-1} - 3 \chi_K)\ .
\end{equation}
One solution is trivial, $\chi_K = 0$, one is unphysical, $\chi_K =
2(1 - X_{K-1})$. The only physically acceptable solution in the
spin-glass phase is $\chi_K = 2(1 - X_{K-1})/3$.

Next we use this physical solution in Eq.~\eqref{eq:chi-final} for $l
= K -1$. We again can factorize this equation for $\chi_{K-1}$ to a
product
\begin{equation}\label{eq:chi-KK-1}
0 = \chi_{K-1}(2 - 2 X_{K-2} - \chi_{K-1})(2 - 2 X_{K-2} - 5 \chi_{K-1})\ .
\end{equation}
The only physical solution is $\chi_{K-1} = 2(1 - X_{K-2})/5$.

We go on with this successive substitutions and derive a generic
physical solution
\begin{equation}\label{eq:chi-final1}
 \chi_{K-l} = \frac 2{2 l + 3}(1 - X_{K-l-1})\ .
\end{equation}
Realizing that $X_0 = 0$ we obtain an explicit solution for all
coefficients $\chi_l, l = 1,2,\ldots,K$.

\section{Properties of the asymptotic solution and the continuous limit}
\label{sec:properties}

After finding an explicit expression for the leading asymptotic terms
of the local overlap susceptibilities we can recover the leading-order
(in some case also the next-to-leading-order asymptotic) coefficients
for all physical quantities of interest. We now use
solution~\eqref{eq:chi-final1} in
Eqs.~\eqref{eq:chi-explicit},~\eqref{eq:Qprime-solution}
and~\eqref{eq:Q-solution} to obtain an explicit asymptotic form of the
order parameters:
\begin{subequations}\label{eq:order-parameters}
\begin{align}\label{eq:chi_i}
  \Delta\chi_l^K &\doteq \frac 2{2 K + 1}\ \theta\ ,\\ \label{eq:m_i}
  m_l^K &\doteq \frac {4 (K - l + 1)}{2 K + 1}\ \theta \\ \label{eq:q}
  q^K &\doteq \frac 1{2 K + 1}\ \theta
\end{align}
and the first two leading orders of the Edwards-Anderson parameter
\begin{align}\label{eq:Qprime}
  Q^K &\doteq \theta + \frac {12 K(K + 1) + 1}{3(2 K +1)^2}\ \theta^2\
  .
\end{align}
\end{subequations}
We returned to the superscript explicitly denoting the dependence of
the order parameters on the number of hierarchies.

First thing we can observe is that both differences $\Delta\chi_l^K$
and $\Delta m_l^K = m_{l-1} - m_l$, $m_0 = 1$, do not depend on the
hierarchy index $l$. We have $\Delta m_l^K = 2 \Delta\chi_l^K = 4/(2 K
+ 1)$. It means that if all solutions with finite hierarchical levels
are unstable we have to choose the number of hierarchies infinite. The
discrete RSB scheme then converges to a continuous theory with either
$\Delta\chi^K$ or $\Delta m^K$ as infinitesimal differentials. The
ratio of the two differentials is in the leading order near the
critical temperature constant, $\Delta\chi^K/\Delta m^K \doteq 1/2$.
This ratio is even independent of the total number of hierarchies $K$.

Before we investigate stability of the solution with $K$ hierarchies
we evaluate other physical quantities of interest with the aid of the
asymptotic form of the order parameters to see how they depend on the
number of hierarchies. With Eqs.~\eqref{eq:order-parameters} we are
able to evaluate the leading deviation beyond the paramagnetic
solution. We start with the thermal local susceptibility $\chi_T$ that
in the spin-glass phase does not obey the single-state Fischer
relation, $\chi_T =\beta(1 - Q)$.\cite{Fischer91} In the RSB solution
we have
\begin{equation}\label{eq:thermal-susceptibility}
\chi_T = \beta \left(1 - Q + \sum_{l=1}^K m_l\Delta\chi_l\right) \doteq 1 -
\frac{\theta^2} {3(2K + 1)^2}\ . 
\end{equation}
We see that in the limit $K\to\infty$ the local susceptibility seems
to be a constant below the critical temperature.

Another physical quantity of interest is the density of internal
energy that behaves near the critical temperature in the following
manner
\begin{multline}\label{eq:internal-energy}
  u = \frac{\partial}{\partial \beta}\left(\beta f\right) = - \frac
  \beta 2 (1 - Q^2) \\ - \frac \beta 2 \sum_{l=1}^K m_l \Delta\chi_l(2
  Q_l -\Delta\chi_l) \doteq -\frac 12 - \frac \theta 2 + \frac
  {\theta^3}3\ .
\end{multline}
The first two terms in the temperature behavior of the internal energy
are from the paramagnetic solution (trivial order parameters). The
third term is fully determined by the SK solution, since it does not
depend on the number of hierarchical levels used.

The last quantity the asymptotic behavior of which we evaluate near
the critical temperature is free energy. We do not need to expand the
free energy directly, since we can use the asymptotic expansion for
the internal energy and the defining equation relating the two
quantities from Eq.~\eqref{eq:internal-energy}. It can be rewritten to
a more suitable form using the expansion parameter $\theta$: $u = f +
(1 -\theta) \partial f/ \partial\theta$. Using this equation and the
asymptotic result from Eq.~\eqref{eq:internal-energy}we easily find
\begin{equation}\label{eq:FE-asymptotic}
f \doteq  -\ln 2  - \frac \theta 2 (2\ln 2 + 1)  - \frac {\theta^2}4
- \frac{\theta^3}{12} + \frac {\theta^4}{24} \ .
\end{equation}
We can see that the leading order asymptotic terms in the order
parameters breaking the replica symmetry contribute to the density of
free energy only in the fifth order of the deviation from the critical
temperature.  This order cannot, however, be determined exactly from
the leading asymptotic form of the stationarity equations for the
order parameters.

We can, nevertheless, improve upon the precision of the asymptotic
expansion of free energy and consequently of internal energy without
increasing the precision of the order parameters. We can apply the
asymptotic expansion in $\theta$ directly to the functional of free
energy. We then are able to expand free energy at the saddle point to
the order $O(\theta^5)$ and internal energy to $O(\theta^4)$. It
appears that the next-to-leading asymptotic contributions to the order
parameters cancel each other in free energy up to the order
$O(\theta^5)$.  Using the asymptotic form of the interacting part of
free energy from Eq.~\eqref{eq:freeEnInt2} in Eq.~\eqref{eq:mf-avfe}
together with the asymptotic solutions for the order parameters we end
up with corrections to the paramagnetic solution of free energy
\begin{equation}\label{eq:FE-better-asymptotic}
\Delta f\doteq\left(\frac{1}{6}\theta^3+\frac{7}{24}
\theta^4+\frac{29}{120}\theta^5\right)-\frac{1}{360} \theta^5
   \left(\frac{1}{K}\right)^4
\end{equation}
and of internal energy
\begin{equation}\label{eq:better-internal-energy}
\Delta u \doteq\left(\frac{1}{2}\theta^2+\frac{5}{6}
\theta^3+\frac{1}{3}\theta^4\right)-\frac{1}{72} \theta^4
   \left(\frac{1}{K}\right)^4 \ .
\end{equation}

The number of hierarchies $K$ is a free parameter in the general
hierarchical solution we analyzed. To make the solution unique we have
to decide what choice of this parameter leads to the exact solution.
Generally, the number of hierarchies in free energy~\eqref{eq:mf-avfe}
is the minimal one for which all stability conditions from
Eq.~\eqref{eq:AT-hierarchical} are satisfied. We have $K + 1$
stability conditions for a given $K$. In the first two leading
asymptotic orders they can be rewritten with the aid of the order
parameters as
\begin{multline}\label{eq:Lambda-discrete}
  \Lambda_l \doteq 1 \\ - \beta^2\left[ 1 - 2 Q + 2 M_l + X_l^2 + 2
    Q_{l+1}X_l + 3 Q_{l+1}^2\right]\ .
\end{multline}
Using the relation between quantities $Q_{l+1}$ and $X_l$ and the
expansion of the order parameters to the leading order and the
Edwards-Anderson parameter to the next-to-leading order we obtain
\begin{multline}\label{eq:Lambda-asymptotic}
  \Lambda_l \doteq - 2\theta(1 - Q)\\ - 2\theta^2\left[(1 - X_l)^2 -
    \frac 13(1 - X_K)^2 + M_l - M_K\right]\ .
 \end{multline}
 It is clear from the solution from Eq.~\eqref{eq:order-parameters}
 that the linear term on the right-hand side of
 Eq.~\eqref{eq:Lambda-asymptotic} vanishes. The first nontrivial
 temperature-dependent term then reads\cite{Note1}
\begin{equation}\label{eq:Lambda-solution}
 \Lambda_l = -\frac  43 \ \frac {\theta^2}{(2 K + 1)^2}\ .
\end{equation}
All the stability conditions collapse in the leading asymptotic order
to a single expression independent of the hierarchical index $l$.
Since the stability parameters $\Lambda_l $ are negative at any $K$,
we conclude that no finite number of hierarchies in
Eq.~\eqref{eq:mf-avfe} is able to produce a physical solution with no
negative values for the stability parameters.  Since
$\lim_{K\to\infty}\Lambda_l = 0$, only the solution with infinite-many
hierarchical levels is physically acceptable.  The RSB solution with
infinite many hierarchies is then marginally stable.

We can see from the asymptotic solution~\eqref{eq:order-parameters}
that the limit of infinite-many hierarchies leads to a continuous
limit with infinitesimal differentials $\Delta\chi_l$ or $\Delta m_l$.
The continuous limit then results in simplifications of the
hierarchical free energy and the corresponding equations of motion.
First, all higher than linear powers of the difference $\Delta\chi_l$
(for the fixed index $l$) vanish from the continuous limit. With this
simplification in mind we can significantly reduce the set of
equations~\eqref{eq:chi-final} for the leading-order coefficients of
the overlap susceptibilities We assign continuous quantities to the
following discrete variables: $X_l\to x$, $m_l \to m(x)$, $X_K \to
x_{max}$. Equation~\eqref{eq:chi-final} then reduces to
\begin{multline}\label{eq:X-continuous}
  3\int_x^{x_{max}}dy(y - x) m(y)\\ = (1 - x)[x_{max}(2 - x_{max})-
  x(2 - x)] \ .
\end{multline}
We know that $x_{max} = 1 - q$. We can now solve
Eq.~\eqref{eq:X-continuous} independently of the discrete
approximations.  We denote $M(x) = \int_0^x dy m(y)$ and $M = M(1)$
and rewrite Eq.~\eqref{eq:X-continuous} to
\begin{multline}\label{eq:M-continuous}
  3(M - M(x)) = x_{max}(2 - x_{max}) - x(2 - x) \\ + 2(1 - x)^2\ .
\end{multline}
From continuity of the function $M(x)$ we obtain $x_{max} = 1$ and
hence $q = 0$. A derivative of this equation with respect to $x$ leads
to an explicit representation
\begin{equation}\label{eq:m-continuous}
m(x) = 2 (1 - x)\ .
\end{equation}
Finally, using the continuous version of
Eq.~\eqref{eq:Qprime-solution} we derive the next-to-leading
asymptotic term of the Edwards-Anderson order parameter $Q' = \int_0^1
dx m(x) = M = 1$.

The solution obtained from the continuous version of the RSB scheme
coincides with the limit of the discrete scheme with infinite number
of hierarchies. All the physical quantities, when infinite-many
hierarchies are needed to reach a stable solution, can hence be
derived directly from the continuous RSB scheme. The only information
we lose in the continuous solution is the rate with which the
continuous limit is approached.

\section{Conclusions}
\label{sec:Conclusions}
 
We studied in this paper the full discrete hierarchy of RSB solutions
of the zero-field SK model in the critical region below the transition
temperature to the spin-glass phase. We expanded the stationarity
equations for all the order parameters to an appropriate order in
$\theta = 1 - T/T_c$ so as to find their leading asymptotic behavior.
We succeeded in solving the resulting equations and obtained an
explicit leading asymptotic behavior of all thermodynamic functions of
the RSB solution with an arbitrary number of hierarchies. The number of
hierarchies in the RSB solution is treated as a free parameter to be
determined from stability conditions. Stability of the RSB scheme with $K$
hierarchies is measured by $K + 1$ numbers $\Lambda_l$, $l=0,1,\ldots,
K$ from Eq.~\eqref{eq:AT-hierarchical}. All these numbers must be
nonnegative in a (marginally) stable solution. We explicitly evaluated
these numbers with the asymptotic solution for the order parameters
and found that any solution with finite-many hierarchies is unstable
with instability of order $\theta^2$. Near the critical temperature
all the stability conditions collapse in the first two leading
asymptotic orders to a single criterion. With the increasing number of
hierarchies the instability parameters decrease and become zero at $K
= \infty$. Hence first the infinite-order RSB is marginally stable
below the critical temperature.

The most important physical issue we addressed in this paper was the
legitimacy of the continuous limit. We proved by explicit calculation
that the discrete RSB scheme indeed converges toward the continuous
limit with $K\to\infty$. We found that near the critical temperature
both the overlap susceptibilities $\chi_l$ and the geometric
parameters $m_l$ determining the way the replica symmetry is broken
are uniformly distributed and their differences $\Delta\chi_l$ and
$\Delta m_l$ become infinitesimal for $K\to\infty$.  Either of these
differences can hence be used as a fundamental differential for the
continuous limit. Since the differences with the same hierarchy index
$l$ appear in the free energy of the continuous limit only linearly,
we loose one set of stationarity equations from the discrete version.
It appeared more natural to choose $dx =
\lim_{K\to\infty}\Delta\chi_K$ as the underlying differential for the
continuous formulation of the RSB solution with infinite-many
hierarchies.  Hence, $\Delta\chi_l$ are no longer variational
parameters in the continuous limit and only $x_{max}$, the maximal
value for the overlap susceptibility is to be determined from the free
energy functional for the continuous limit of the RSB scheme.

The asset of the present construction, however, is not only in that it
provides a proof of exactness of the continuous limit for the
zero-field SK model near the critical temperature.  This conclusion
was actually believed to be correct already from the existing less
accurate treatments.  With our construction we set up a general scheme
how to solve the RSB equations with an arbitrary number of hierarchies
near the critical point practically for any mean-field spin-glass
model. We applied this construction to the case of zero magnetic field
in the SK model. But the same scheme can be applied also in the
presence of a magnetic field and expand the hierarchical solution
around the de Almeida-Thouless line. In this case, however, the
geometric order parameters are no longer small. We can as well apply
the expansion scheme to other mean-field spin-glass models. Of
particular interest are those where one expects that in certain
parameter regions the RSB solutions with a few hierarchies (1RSB) are
stable, such as the Potts spin glass. The question that has not yet
been answered is whether the instability of the 1RSB solution there
leads already to the continuous limit or not. The scheme presented
here can conveniently be used to address this problem.

\begin{acknowledgments}
  Research on this problem was carried out within a project
  AVOZ10100520 of the Academy of Sciences of the Czech Republic and
  supported in part by Grant No. IAA1010307 of the Grant Agency of the
  Academy of Sciences of the Czech Republic.
\end{acknowledgments}

\appendix
\section{Expansion coefficients for  partition sum near the critical temperature}
\label{sec:Expnansion-partion-sum}

Here we list the expansion coefficients needed for the asymptotic
solution near the critical temperature. We evaluate the coefficients
used for the $l$th partition sum in Eq.~\eqref{eq:Z_l-expansion}. We
used MATHEMATICA to derive the recursive relations from which we then
obtained
\begin{align}\label{eq:Recursive-coeffs1}
a^{(l)}_{0i} &= a_{0i}\\ %
a^{(l)}_{20} &= a_{02}\sum_{i=1}^l c_i^2\\%
a^{(l)}_{22} &= 6
a_{04}\sum_{i=1}^l c_i^2 + 2 a_{02}^2 \sum_{i=1}^l (m_i
- 1) c_i^2  \\  %
a^{(l)}_{24} &= 15 a_{06} \sum_{i=1}^l c_i^2 + 2 a_{02}
(4 a_{04} - a_{02}^2) \sum_{i=1}^l (m_i - 1) c_i^2 %
\end{align}
\begin{widetext}
\begin{align}\label{eq:Recursive-coeffs2}
  a^{(l)}_{40} &= 3 a_{04} \sum_{i=1}^l c_i^2 \left[ 2
    \sum_{j=1}^{i-1} c_j^2 + c_i^2\right] + a_{02}^2 \sum_{i=1}^l
  c_i^2 \left[ 2 \sum_{j=1}^{i-1} (m_j - 1)c_j^2 + (m_i -
    1)c_i^2\right]
\end{align} %
\begin{align}
  a^{(l)}_{42} &= 45 a_{06} \sum_{i=1}^l c_i^2 \left[ 2
    \sum_{j=1}^{i-1} c_j^2 + c_i^2\right] + 12 a_{02} a_{04}
  \sum_{i=1}^l c_i^2 \left[ (m_i - 1) \left(3 c_i^2 + 2
      \sum_{j=1}^{i-1} c_j^2\right) + 4 \sum_{j=1}^{i-1}
(m_j - 1) c_j^2\right] \nonumber\\ %
&\qquad + a_{02}^3 \sum_{i=1}^l c_i^2 \left\{ (m_i - 1) \left[ (4m_i
    -11) c_i^2 + 2 \sum_{j=1}^{i-1}(4m_j - 5) c_j^2\right] - 12
  \sum_{j=1}^{i-1}
(m_j - 1) c_j^2 \right\}  %
\end{align}
\begin{align}
  a^{(l)}_{60} &= 15 a_{06} \sum_{i=1}^l c_i^2 \left[ c_i^4 + 3 c_i^2
    \sum_{j=1}^{i-1} c_j^2 + 3 \sum_{j=1}^{i-1} c_j^2 \left(2
      \sum_{k=1}^{j-1} c_k^2 + c_j^2\right)\right] + 12 a_{02} a_{04}
  \sum_{i=1}^l c_i^2 \left\{ (m_i - 1) c_i^2 \left( c_i^2 +
\sum_{j=1}^{i-1} c_j^2 \right)\right. \nonumber \\ %
& \left. \qquad + 2 c_i^2 \sum_{j=1}^{i-1}(m_j - 1) c_j^2 +
  \sum_{j=1}^{i-1} c_j^2 \left[ (m_j - 1)\left( c_j^2 + 2
      \sum_{k=1}^{j-1}c_k^2\right) + 4 \sum_{k=1}^{j-1}(m_k - 1) c_k^2
  \right]
\right\}\nonumber\\ %
& \quad + a_{02}^3 \sum_{i=1}^l c_i^2 \left\{\frac 13 (m_i - 1) c_i^2
  \left[ (4m_i -11) c_i^2 + 3 \sum_{j=1}^{i-1} (4m_j - 5)c_j^2 \right]
  - 6
c_i^2 \sum_{j=1}^{i-1} (m_j - 1)c_j^2\right. \nonumber\\ %
& \left.\qquad - 12 \sum_{j=1}^{i-1}c_j^2 \sum_{k=1}^{j-1}(m_k - 1)
  c_k^2 + \sum_{j=1}^{i-1} (m_j - 1)c_j^2 \left[ (4m_j - 11) c_j^2 + 2
    \sum_{k=1}^{j-1}(4m_k - 5) c_k^2 \right]\right\}
\end{align}
\end{widetext}
The initial coefficients are determined by the hyperbolic cosine in
$Z_0$ and read $a_{00} = 1$, $a_{02} = 1/2$, $a_{04} = 1/24$, and
$a_{06} = 1/720$.

\section{Direct asymptotic expansion of free energy}
\label{sec:Expansion-free-energy}

Instead of expanding the stationarity equations we can expand directly
free energy near the critical point. The advantage of such an
expansion is a possibility to go to higher orders of the small
expansion parameter more easier than in the case of stationarity
equations. The drawback of a direct expansion of free energy is that
when building up the corresponding stationarity equations we have to
vary w.r.t. small parameters and we change the order of the expansion
to which we keep all terms exact. It then may happen that different
stationarity equations are evaluated not in the same order of the
expansion parameter.
 
The expansion of free energy is achieved in the same manner in which
we treated the partition sum with the expansion coefficients from
Appendix A.  Using the program MATHEMATICA we were able to go to the
fifth order in $\theta$. In this expansion we, however took into
consideration the fact that the geometric parameters $m_l\propto
\theta$ and neglected all combinations of the order parameters of
order $O(\theta^6)$.  It appears that the next-to-leading order
contributions to the order parameters actually contribute to free
energy nontrivially first in the order $\theta^6$. It is not
manifestly evident but an explicit expansion discloses this feature.
If we denote $\mathcal{G}_K= \int_{-\infty}^\infty \mathcal{D}\eta \ln
Z_K$ we obtain an exact asymptotic for arbitrary number of
hierarchical levels
\begin{widetext}
\begin{align}\label{eq:freeEnInt2}
  \mathcal{G}^{K}_{5} &\equiv \sum _{l=1}^K \left\{\frac{31}{15}
    \Delta \chi _l^5 \beta ^{10}+\left(\frac{31 q \beta ^2}{3}+\frac{4
        m_l}{3}-\frac{17}{24}\right) \Delta \chi _l^4 \beta
    ^8+\left(\frac{62 q^2 \beta ^4}{3}-\frac{17 q \beta
        ^2}{6}+\frac{16}{3} q m_l \beta
      ^2+\frac{m_l^2}{6}-\frac{m_l}{2}+\frac{1}{3}\right) \Delta \chi
    _l^3 \beta ^6 \right.
   \nonumber \\ %
   & \left. +\left(\frac{62 q^3 \beta ^6}{3}-\frac{17 q^2 \beta
         ^4}{4}+\frac{27}{4} q^2 m_l \beta ^4+\frac{1}{2} q m_l^2
       \beta ^2+q \beta ^2-\frac{3}{2} q m_l \beta
       ^2+\frac{m_l}{4}-\frac{1}{4}\right) \Delta \chi _l^2 \beta
     ^4\right.
   \nonumber \\ %
   & \left.+\left(\frac{1}{2} \left(\frac{62 q^4 \beta ^8}{3}-\frac{17
           q^3 \beta ^6}{3}+2 q^2 \beta ^4-q \beta
         ^2\right)+\frac{1}{2} \left(\frac{17 q^3 \beta ^6}{3}-2 q^2
         \beta ^4+q \beta ^2\right) m_l+\frac{1}{2}\right) \Delta \chi
     _l \beta ^2\right\}
   \nonumber \\ %
   & +\sum _{l=1}^K \sum _{j=l+1}^K \left\{\frac{62}{3} \left(\Delta
       \chi _l^2 \Delta \chi _j^3+\Delta \chi _l^3 \Delta \chi
       _j^2\right) \beta ^{10}+\frac{31}{3} \left(\Delta \chi _l
       \Delta \chi _j^4+\Delta \chi _l^4 \Delta \chi _j\right) \beta
     ^{10}+\left(\frac{124 q \beta ^2}{3}+\frac{16
         m_l}{3}-\frac{17}{6}\right) \Delta \chi _j \Delta \chi _l^3
     \beta ^8\right.
   \nonumber \\ %
   & \left.+\left(62 q \beta ^2+\frac{5 m_j}{4}+\frac{27
         m_l}{4}-\frac{17}{4}\right) \Delta \chi _j^2 \Delta \chi _l^2
     \beta ^8+\left(\frac{124 q \beta ^2}{3}+\frac{5 m_j}{2}+\frac{17
         m_l}{6}-\frac{17}{6}\right) \Delta \chi _j^3 \Delta \chi _l
     \beta ^8\right.
   \nonumber \\ %
   & \left.+\left(62 q^2 \beta ^4-\frac{17 q \beta ^2}{2}+\frac{5}{2}
       q m_j \beta ^2+\frac{27}{2} q m_l \beta
       ^2+\frac{m_l^2}{2}-\frac{3 m_l}{2}+1\right) \Delta \chi _j
     \Delta \chi _l^2 \beta ^6\right.
   \nonumber \\ %
   & \left.+\left(62 q^2 \beta ^4-\frac{17 q \beta ^2}{2}+\frac{15}{2}
       q m_j \beta ^2+\frac{17}{2} q m_l \beta
       ^2-\frac{m_j}{2}+\frac{m_j m_l}{2}-m_l+1\right) \Delta \chi
     _j^2 \Delta \chi _l \beta ^6\right.
   \nonumber \\ %
   & \left.+\left(\frac{124 q^3 \beta ^6}{3}-\frac{17 q^2 \beta
         ^4}{2}+5 q^2 m_j \beta ^4+\frac{17}{2} q^2 m_l \beta ^4+2 q
       \beta ^2-q m_j \beta ^2-2 q m_l \beta ^2+q m_j m_l \beta
       ^2+\frac{m_l}{2}-\frac{1}{2}\right) \Delta \chi _j \Delta \chi
     _l \beta ^4\right\}
   \nonumber \\ %
   & +\sum _{l=1}^K \sum _{j=l+1}^K \sum _{i=j+1}^K \left\{62
     \left(\Delta \chi _i \Delta \chi _l^2 \Delta \chi _j^2+\Delta
       \chi _i^2 \Delta \chi _l \Delta \chi _j^2+\Delta \chi _i^2
       \Delta \chi _l^2 \Delta \chi _j\right) \beta ^{10}\right.
   \nonumber \\ %
   & \left.+\frac{124}{3} \left(\Delta \chi _j \Delta \chi _l \Delta
       \chi _i^3+\Delta \chi _j \Delta \chi _l^3 \Delta \chi _i+\Delta
       \chi _j^3 \Delta \chi _l \Delta \chi _i\right) \beta
     ^{10}+\left(124 q \beta ^2+\frac{5 m_j}{2}+\frac{27
         m_l}{2}-\frac{17}{2}\right) \Delta \chi _i \Delta \chi _j
     \Delta \chi _l^2 \beta ^8\right.
   \nonumber \\ %
   & \left.+\left(124 q \beta ^2+\frac{15 m_j}{2}+\frac{17
         m_l}{2}-\frac{17}{2}\right) \Delta \chi _i \Delta \chi _j^2
     \Delta \chi _l \beta ^8+\left(124 q \beta ^2+\frac{5 m_i}{2}+5
       m_j+\frac{17 m_l}{2}-\frac{17}{2}\right) \Delta \chi _i^2
     \Delta \chi _j \Delta \chi _l \beta ^8\right.
   \nonumber \\ %
   & \left.+\left(124 q^2 \beta ^4-17 q \beta ^2+5 q m_i \beta ^2+10 q
       m_j \beta ^2+17 q m_l \beta ^2-m_j+m_j m_l-2 m_l+2\right)
     \Delta \chi _i \Delta \chi _j \Delta \chi _l \beta ^6\right\}
   \nonumber \\ %
   & +\sum _{l=1}^K \sum _{j=l+1}^K \sum _{i=j+1}^K \sum _{k=i+1}^K
   \left\{124 \left(\Delta \chi _j \Delta \chi _k \Delta \chi _l
       \Delta \chi _i^2+\Delta \chi _j \Delta \chi _k \Delta \chi _l^2
       \Delta \chi _i+\Delta \chi _j \Delta \chi _k^2 \Delta \chi _l
       \Delta \chi _i+\Delta \chi _j^2 \Delta \chi _k \Delta \chi _l
       \Delta \chi _i\right) \beta ^{10}\right.
   \nonumber \\ %
   & \left.+\left(248 q \beta ^2+5 m_i+10 m_j+17 m_l-17\right) \Delta
     \chi _i \Delta \chi _j \Delta \chi _k \Delta \chi _l \beta
     ^8\right\}
   \nonumber \\ %
   & +\sum _{l=1}^K \sum _{j=l+1}^K \sum _{i=j+1}^K \sum _{k=i+1}^K
   \sum _{p=k+1}^K 248 \beta ^{10} \Delta \chi _i \Delta \chi _j
   \Delta \chi _k \Delta \chi _l \Delta \chi _p \ .
\end{align}
\end{widetext}
Although the interacting part of free energy $\mathcal{G}_K$ is of one
order higher than the partition sum we used in the stationarity
equations, we are unable to evaluate the next-to-leading order of the
asymptotic solutions for the order parameters. To calculate the second
asymptotic coefficients of the order parameters, in particular the
geometric ones, one had to know the complete form of free energy to
order $O(\theta^6)$.

\end{document}